# Monthly Rural-Urban Scaling of Road Accidents in England, Wales and Scotland (2019-2023)

Isabel Copsey[a], Quentin Hanley[a,b], Jack Sutton[a*]

[a]*University of Derby, Data Science Research Centre, College of Science and Engineering, United Kingdom*

[b]*GH and Q Services Limited, West Studios, Sheffield Road, Chesterfield S41 7LL, UK*

[*]**Corresponding author**: J.Sutton (email: <j.sutton@derby.ac.uk>)

# Abstract

Road traffic accidents remain a major public health challenge worldwide, with urbanisation and population density identified as key factors influencing risk. This study analyses monthly accident data from 2009 to 2023 across 632 parliamentary constituencies in England, Wales, and Scotland, using an area-normalised approach based on population density. Segmented power law models consistently identified breakpoints separating sublinear rural from superlinear urban scaling behaviours. Seasonal variation in scaling exponents was pronounced in rural regions but less evident in urban ones. Fourier-based cross-spectral analysis of yearly cycles revealed systematic phase shifts: rural exponents lagged pre-exponential factors by 4.5 months, while urban exponents were 2.7 months out of phase, producing a 5.3 month shift between rural and urban exponents. These findings highlight the importance of pre-exponentials—defined as the expected density of accidents at unit population density—as comparable descaled metrics, revealing both long-term national declines and recurring seasonal peaks. Notably, the phase offsets suggest structurally distinct causes of rural and urban accident risk, with urban regions exhibiting increasing acceleration in accident scaling, potentially linked to growth in vehicle numbers, size, and weight. Residuals, modelled with the Type I Generalised Logistic Distribution (GLD), captured skewness and heterogeneity more effectively than normal assumptions. Geospatial mapping highlighted persistent urban hotspots alongside rural and coastal constituencies with systematically lower accident densities than predicted. Together, these findings advance understanding of how density and urbanisation shape accident risk and provide evidence to support more targeted road safety interventions and policy planning.

# 1 Introduction

In 2023, it was reported that road traffic accidents were responsible for 1.19 million deaths worldwide each year, demonstrating the urgent need to improve road safety [1]. Reflecting this global challenge, in 2024 Great Britain recorded an estimated 128,375 road casualties of all severities. Within this total, 29,537 were either killed or seriously injured [2]. These figures highlight the continuing scale of the problem and demonstrate that road traffic incidents remain a major public health and policy challenge.

One of the most significant factors contributing to this challenge is increasing urbanisation, which brings higher traffic volumes, congestion, and greater exposure to risk. Today, more than 4 billion people, over half of the world's population, live in urban areas [3]. This share has risen sharply in recent decades and is projected to keep growing, highlighting the importance of understanding how urbanisation influences road safety outcomes.

A study in England and Wales illustrates this relationship clearly [4]. Urban areas experience more minor and serious accidents, while fatal accidents are more common in rural settings. As city populations grow, the overall number of accidents tends to increase faster than population size, particularly for less severe incidents. Yet, the likelihood that any given accident is fatal or serious decreases in larger urban areas, with minor accidents instead accounting for most of the growth. These patterns point to a complex but important connection between urbanisation, population size, and road safety. Similar investigations and the need to understand road accidents have also been conducted in other countries, including Lebanon [5] and Belgium [6].

In this study, we analyse monthly reported road accident data across rural and urban regions, defined by parliamentary constituencies, over the period 2009–2023. An area normalised approach based on population density, rather than absolute population, is employed to capture the full continuum of environments. Segmented power law models are used to examine rural and urban scaling regimes, with pre-exponential factors explicitly considered as descaled metrics reflecting expected accident density at unit population density. To investigate seasonal dynamics, Fourier-based cross-spectral analysis is applied to the yearly cycles of pre-exponentials and exponents, revealing systematic phase shifts between rural and urban behaviours. Residuals from the segmented fits are modelled throughout the study period using Generalised Logistic Distributions, which capture skewness and heterogeneity more effectively than normal assumptions. Finally, geomapping identifies constituencies with systematically high or low accident densities relative to model predictions, highlighting persistent urban hotspots as well as rural and coastal deviations.

# 2 Methodology

## 2.1 Theory

Traditional urban scaling models [7–11] assume that total quantities of interest $Y$ (e.g. Road Accidents) scale with population $P$ according to a power law:

$$Y = Y_0 P^\beta \qquad (1)$$

where $Y_0$ is the pre-exponential factor and $\beta$ is the scaling exponent.

In this study, we follow a recently proposed adjustment of Equation 1 to account for rural areas by normalising by area $(A)$, converting total values to densities such that $D_P = P/A$ and $D_Y = Y/A$ [12–18]. Accordingly, the scaling model and its corresponding logarithmic version become:

$$D_Y = Y_0 D_P^\beta \qquad (2)$$

$$log(D_Y) = log(Y_0) + \beta log(D_P)$$

where $log(Y_0)$ is the pre-exponential intercept and $\beta$ becomes the density scaling exponent describing how road accident density responds to changes in population density. When $\beta < 1$, the relationship exhibits sublinear scaling, indicating that road accident density has diminishing returns with population density. When $\beta = 1$, scaling is linear, meaning accident density grows proportionally with population density. When $\beta > 1$, scaling is superlinear, indicating increasing road accident with higher population density.

Rural and urban areas often exhibit different scaling behaviours. To capture these, we use a piecewise (segmented) power law model with a population density breakpoint ($D_P^*$):

$$\begin{array}{ll} \alpha_1 + \beta_1 log(D_P) + \varepsilon, & if\ D_P \leq D_P^*\quad \text{(rural regime)} \\ \alpha_2 + \beta_2 log(D_P) + \varepsilon, & if\ D_P > D_P^*\quad \text{(urban regime)} \end{array} \qquad (3)$$

where $D_P^*$ is the estimated threshold population density separating rural and urban scaling regimes, $\beta_1$ represents rural scaling behaviour, $\beta_2$ represents urban scaling behaviour and $(\alpha_1, \alpha_2)$ are intercepts for the respective segments. The breakpoint $D_P^*$ is estimated using Davies test and model comparisons between single and segmented models is checked using AIC and BIC scores. Lastly, the residuals $\varepsilon$ represent the difference between the actual observations and their expected values. Larger residuals indicate greater deviations from the model. Positive residuals occur when observations exceed expectations, while negative residuals indicate values below expectations relative to the power law and population density.

Residuals are obtained from the best fitted scaling model using power laws from (Equations 2-3). These are examined using the Type I Generalised Logistic Distribution (GLD):

$$f(x; \theta, \sigma, \alpha) = \frac{\alpha}{\sigma} \frac{exp\left\{\frac{-(x-\theta)}{\sigma}\right\}}{\left[1 + exp\left\{\frac{-(x-\theta)}{\sigma}\right\}\right]^{\alpha+1}} \qquad (4)$$

where $\theta$ is the location, $\sigma$ is the scale and $\alpha$ is the shape parameters where $\theta \in \mathbb{R}$, $\sigma, \alpha > 0$ and $-\infty < x < +\infty$. The three sample moments corresponding to the mean, variance and skew are as follows:

$$E(X) = \theta + \sigma\{\psi(\alpha) - \psi(1)\} \qquad (5)$$
$$\text{Var}(X) = \sigma^2\{\psi'(1) + \psi'(\alpha)\}$$
$$\text{Skew}(X) = \frac{\psi'(\alpha) - \psi'(1)}{\{\psi'(\alpha) + \psi'(1)\}^{\frac{3}{2}}}$$

where $\psi(\cdot)$ is the digamma function, and $\psi'(\cdot)$ and $\psi''(\cdot)$ are its first and second derivatives, respectively. These moments are used to examine residual variance and skew.

## 2.2 Data

England and Wales population data was taken from the Office for National Statistics (ONS) mid-year constituency population estimates for 2019. Scottish population data was taken from the National Records of Scotland (NRS) mid-year constituency population estimates for 2019. Parliamentary constituency area data was taken from the ONS Geoportal, measured in 2019. Road accident data was collected from UKCrimeStats, a platform of the Economic Policy Centre, which sources its data from the Department for Transport via the UK Government's road safety statistics. Shapefiles were obtained from UK Data Service licensed under the UK Open Government Licence to produce maps.

Population, land area and monthly road accident data were aligned using parliamentary constituencies as defined in the 2019 UK general election and provided in the supplementary material (S1 Dataset). Yearly car characteristics such as width (mm), length (mm), height (mm) and weight (kg) were obtain from a car database and provided in the supplementary material (S2 Dataset). Zero values in the data have been treated as null values; this is particularly prominent in the time period of the COVID-19 pandemic. The treatment of zero values in urban scaling has been discussed extensively in the literature [19] due to the logarithmic transformation; however, this was not an issue in this study since the absence of data was limited and had little impact on the results. The highest number of missing cases occurred in April 2020 due to the COVID-19 restrictions, when 18 parliamentary constituencies (out of 632) reported no road accidents.

## 2.3 Software

Modelling and data analysis were conducted in R (version 4.5.1) [20] and RStudio (version 2025.05.1). Data loading, writing, and formatting were carried out using the rio (0.5.27) [21] and xlsx (0.6.5) [22] packages. Maps of England, Wales and Scotland were generated and visualised using the sf (1.0-21) [23], gplots (3.1.3) [24], ggplot2 (3.5.2) [25], and RColorBrewer (1.1-3) [26] packages. Segmented power law models were fitted using the segmented (2.1-4) [27] package. Residuals were modelled using the Generalised Logistic Distribution (GLD) using the glogis package (1.0-2) [28]. The phase shifts between pre-exponential, rural exponent, and urban exponent were computed following removal of any linear trends by Fourier transforming 14 years of monthly values and using the cross-power spectra to estimate the phase shift. The R script to perform the analysis in this study is available in the supplementary material (S1 Code).

# 3 Results

## 3.1 Regions and Road Accidents

The analysis covers 632 parliamentary constituencies across England, Wales, and Scotland. Populations range from 26,720 in Na h-Eileanan an Iar (Scotland) to 1,888,634 in West Ham (England). Land areas range from 738.04 hectares in Islington North (England) to 1,232,699.84 hectares in Ross, Skye and Lochaber (Scotland). Correspondingly, population density varies from 0.05 persons per hectare in Ross, Skye and Lochaber (Scotland) to 170.79 persons per hectare in Westminster North (England).

The study explored monthly road accident densities over an extended time series from January 2009 to December 2022. Figures 1 and 2 show examples of accident density maps in September 2022, where white indicates parliamentary constituencies with high road accident density and dark blue indicates constituencies with low densities. Across the UK, a number of constituencies consistently emerge as urban hotspots for road accidents. These are areas where population density is high, congestion is greater, and opportunities for accidents increase. The pattern remains relatively consistent over the study period. All monthly maps, both at the national scale and by individual regions, are provided in the Supplementary Materials (S1 Figure).

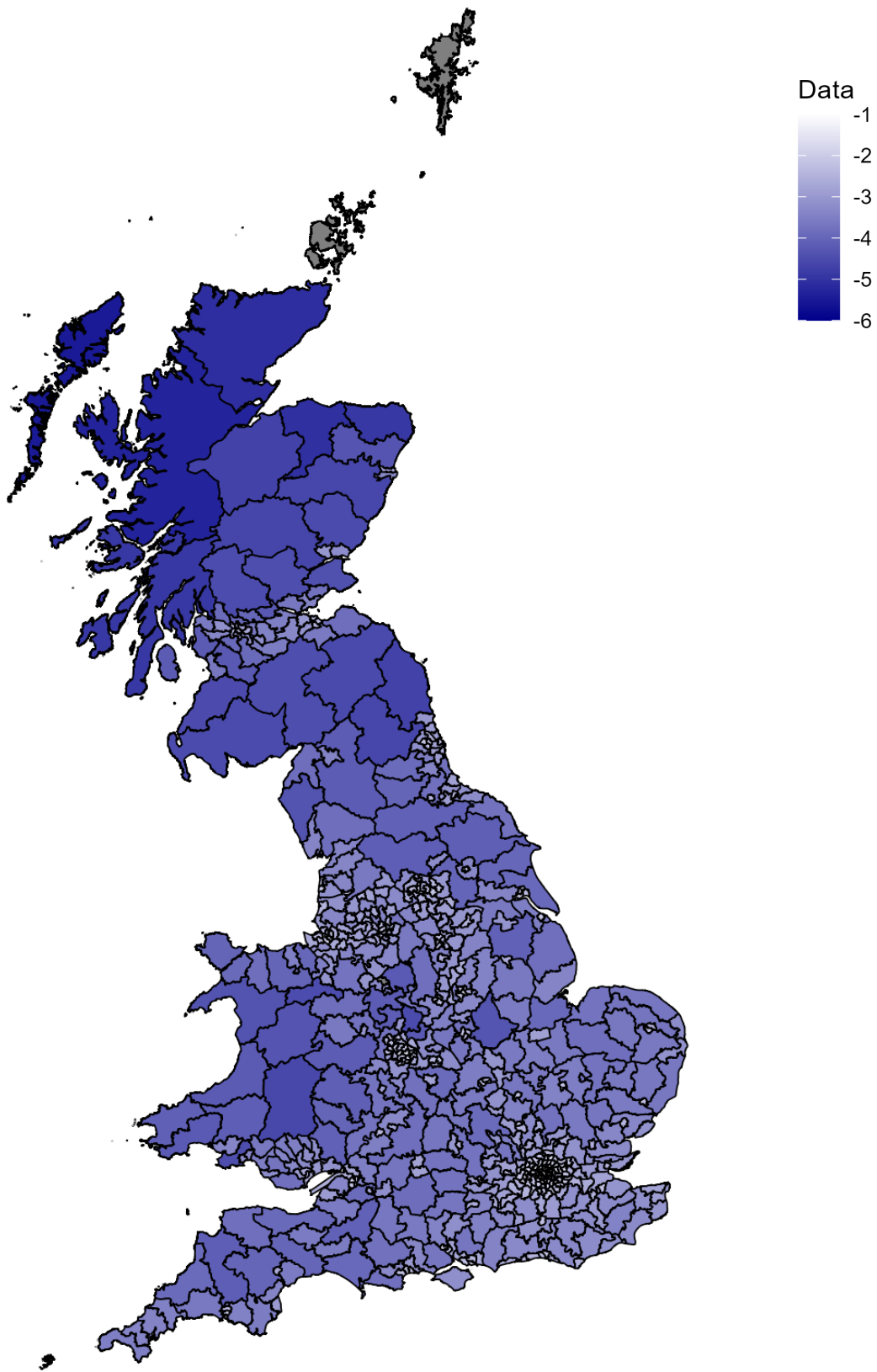

**Figure 1.** Map of road accident densities in England, Wales, and Scotland in September 2022, using parliamentary constituencies as boundaries. Colours range from dark blue to white, with white indicating constituencies with high road accident

densities and dark blue indicating low densities. Grey areas represent regions with no recorded road accidents for that month.

### East Midlands
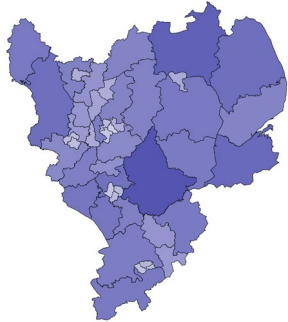

### East of England
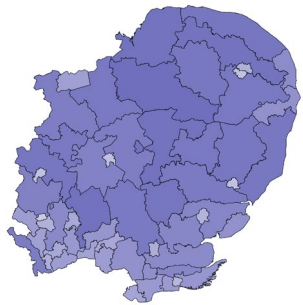

### London
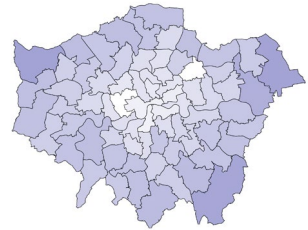

### North East
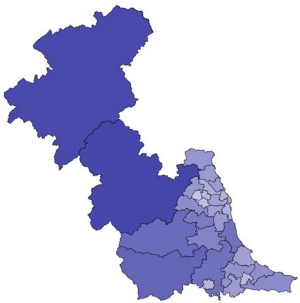

### North West
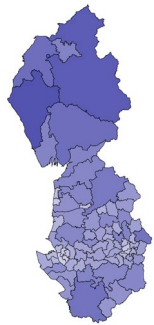

### South East
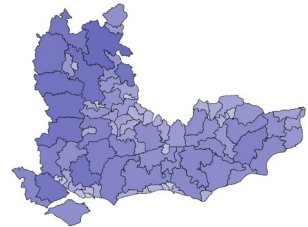

### South West
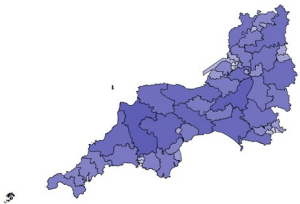

### West Midlands
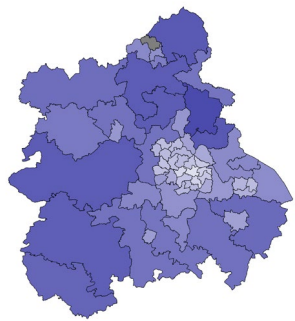

### Yorkshire and The Humber
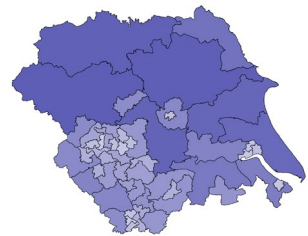

### Wales
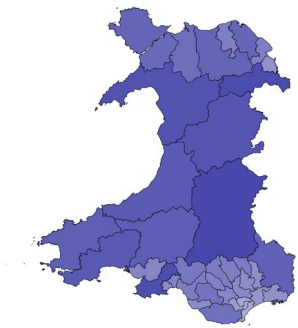

### Scotland
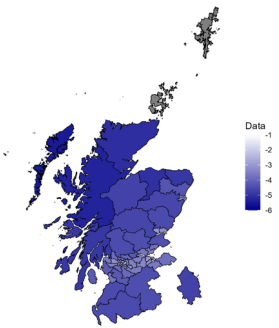

**Figure 2.** Zoomed-in maps of regions from Figure 1 for ease of view, including the East Midlands, East of England, London, North East, North West, South East, South West, West Midlands, Yorkshire and The Humber, Wales, and Scotland. Colours range from dark blue to white, with white indicating constituencies with high road accident densities and dark blue indicating low densities. Grey areas represent regions with no recorded road accidents for that month.

## 3.2 Scaling

Scaling models were applied, with segmentation providing the best fit consistently across the monthly data over the entire studied period. The segmented model held even through the COVID-19 pandemic, demonstrating that it provides a robust fit across a range of population behaviours (Figure 3). A breakpoint was required throughout, with an average density value of 1.318 [CI: 1.178–1.458], equivalent to roughly 20 persons per hectare, separating the lower and upper segments (see Figure 3). Notably, the segmentation remains consistent, with no noticeable deviations among parliamentary constituencies from the power law. This pattern aligns with findings in other contexts, including crime, property transactions, health, and demographics [14–18]. Complete monthly scaling plots for the entire study period are provided in the Supplementary Materials (Figure S2).

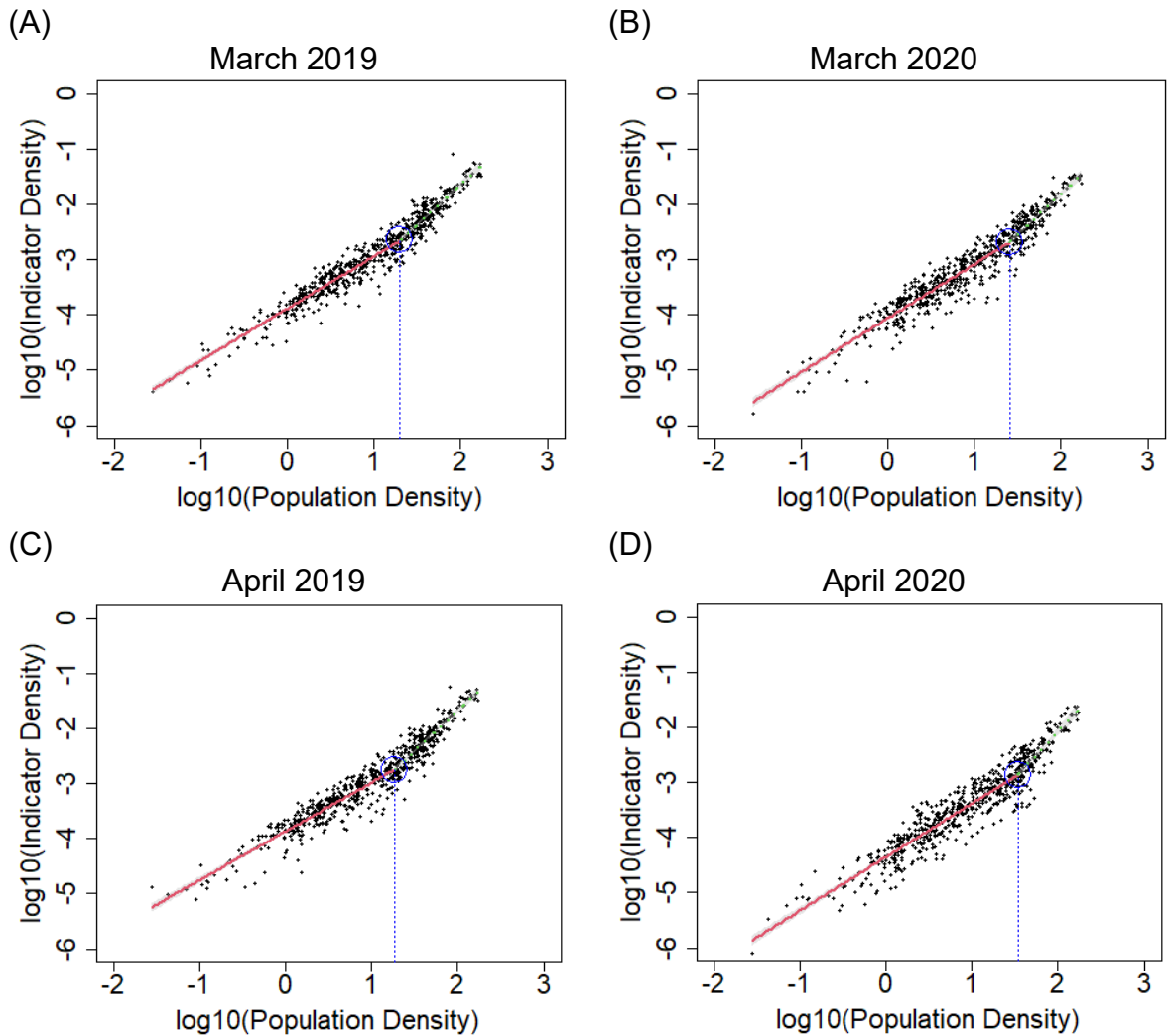

**Figure 3.** Relationship between population density and road accident density across parliamentary constituencies (log–log scale) in (A) March 2019, (B) March 2020, (C) April 2019, and (D) April 2020. The segmented regression fit (solid red–green line) is separated by a breakpoint (blue circle and vertical dashed line) indicating different rural-urban scaling behaviours.

## 3.3 Monthly pre-exponential factors and scaling exponents

To our knowledge, there is no prior interpretation of pre-exponential factors in the scaling field, and this study provides the first opportunity to do so. The logarithmic pre-exponential factors ranged from –4.35 in April 2020 to –3.62 in August 2010 (Figure 4; Panel A). Across the study period, there was a gradual decline, accompanied by a clear seasonal pattern in which peaks were typically observed during July and August. The overall decline in the pre-exponential factor broadly follows the UK trend toward reduction in road accidents over time. The seasonal variation suggests that road accidents were more frequent during summer, reflecting periods of heavier traffic flows and increased travel, particularly during holiday months. Two exceptional drops were observed, in April 2020 and February 2021. These months correspond to the first UK national lockdown, announced on 23 March 2020, and the third lockdown, announced on 6 January 2021, during the COVID-19 pandemic. During these periods, widespread restrictions led to the closure of large parts of the economy, with many people working from home and road transport and national mobility being markedly reduced.

Segmentation was universal, with a consistent breakpoint (Figures 4; Panels B-C) separating two fundamentally different rural–urban scaling regimes. The lower scaling exponent (red dots in Figure 4; Panel B) was almost entirely sublinear (exceptions in November 2019, March 2021, and November 2022). In contrast, the upper scaling exponent was almost entirely superlinear (exception in January 2009). The consistent superlinearity above the breakpoint demonstrates that road accidents in urban areas rise more than proportionally with population density, reflecting the heightened risks associated with crowded urban environments. The gradual increase observed over the extended period further implies that not only do urban environments concentrate accidents, but that this concentration has intensified over time. The rural lower segment displayed clear seasonality, with marked declines in the summer months, whereas seasonality in the urban upper segment was less pronounced.

What is striking in the data is the phase shift between the pre-exponential factors, and the rural and urban exponents. The rural exponent was shifted 4.5 months from the pre-exponential. In other words, if a maximum occurs in the pre-exponential factors in July to August, the low-exponent will peak in late November to early January. In contrast, the high exponent preceded the pre-exponential by 2.7 months The shift between rural and urban exponents was 5.3 months. These results indicate the importance of understanding the pre-exponential in addition to the exponents. Here, the pre-exponential is the density of road accidents expected when the population density is one. More generally, a pre-exponential factor is the expected indicator density at population density = 1. These pre-exponentials should be comparable across different indicators as descaled comparable metrics. Here, the decline in the pre-exponential metric follows the decline in road accidents. The scaling exponents show that the number of accidents is out of phase between rural and urban regions and hence, so are the causes. It also highlights a structural change where urban regions are showing increasing acceleration of road accidents (high exponents). The reasons for this are unclear but may be related to increasing numbers of vehicles and their larger size and weight all of which increased over the period of study in the UK (Figure S3).

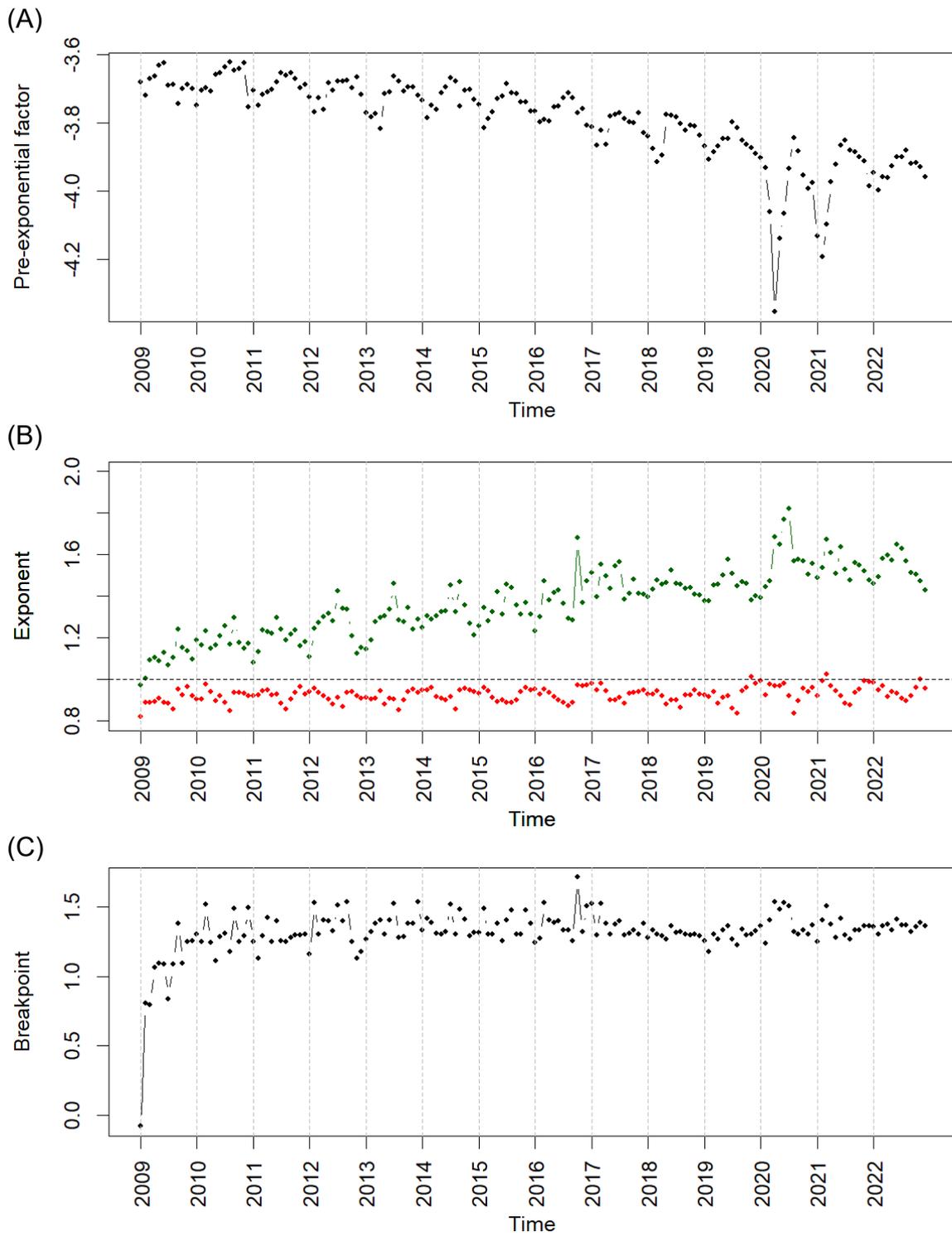

**Figure 4.** Monthly time series of (A) pre-exponential factors, (B) rural–urban scaling exponents, and (C) breakpoints. In (B), red dots represent scaling exponents below the breakpoint, while green dots represent scaling exponents above the breakpoint. The horizontal black dashed line in panel (B) indicates linear scaling.

## 3.4 Monthly residual variance and skew

Residual variance was broadly consistent across the study period, ranging from a low of 0.061 in September 2021 to a peak of 0.110 in April 2020 (Figure 5; Panel A). This exceptional peak corresponds to the COVID-19 pandemic, which caused pronounced national heterogeneity during a time of profoundly altered human behaviour, with strict restrictions and enforce reduced mobility.

In contrast, residual skew displayed less consistency, decreasing by roughly a factor of three across the study period, from a peak of 0.965 in September 2010 (Figure 5; Panel C) to 0.307 in September 2022 (Figure 5; Panel D). Because the skew remained below 1, the residuals were consistently negatively skewed, indicating that some regions fell below the power law expectation indicating few road accident densities. The decline in skew suggests that this left tail became more pronounced over time, with certain regions deviating further from expectations. Monthly residual histograms for the full study period are provided in the Supplementary Materials (Figure S4).

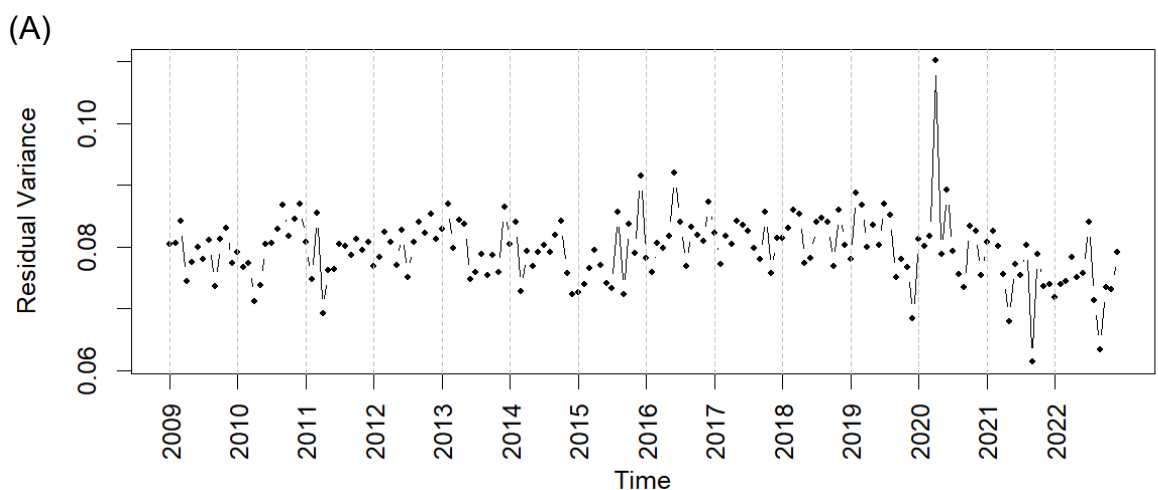

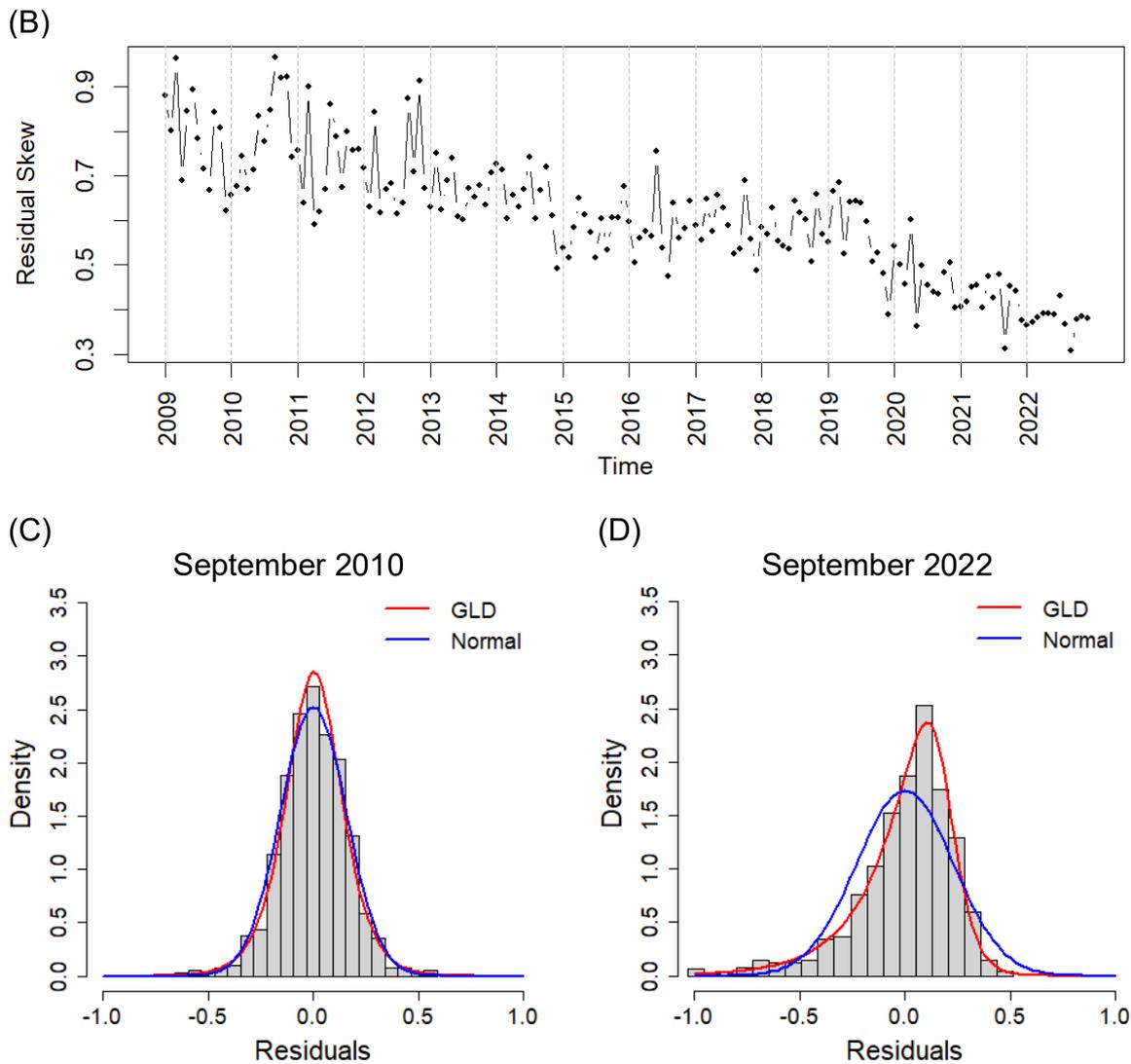

**Figure 5.** Monthly time series of (A) residual variance and (B) residual skew. Examples of residual skew are shown in panels (C–D), where the red solid line represents the GLD and the blue solid line represents the normal distribution. The latter is a common statistical assumption that is clearly violated in this context.

To identify regions that deviate from the power-law expectation, we mapped the residuals—the differences between observed road accident densities and those predicted by road density (see Figure 6). Areas with accident densities above expectation are shaded from orange (low positive deviation) to red (high positive deviation), while those below expectation are shown from light blue (low negative deviation) to dark blue (high negative deviation).

Although urban hotspots with high accident densities are visible in Figures 1–2, these patterns are generally consistent with higher population densities and therefore align with the model expectation. In contrast, the residual map (Figure 6) highlights regions where accident densities diverge from population-based expectations. These deviations are most pronounced in rural and many coastal areas, where accident densities tended to fall below the expected values. Monthly residual maps for the full study period are provided in the Supplementary Materials (Figure S5).

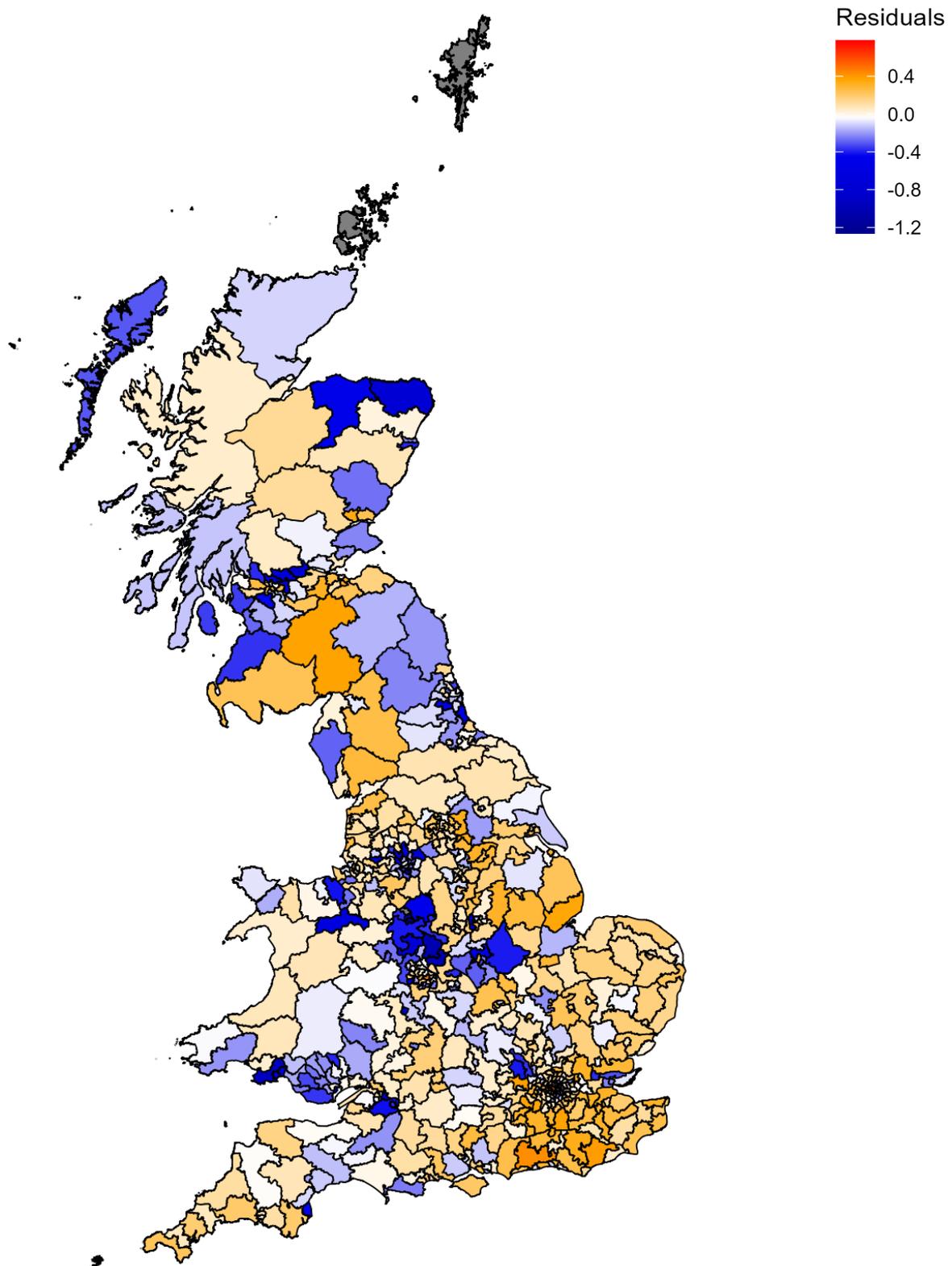

**Figure 6.** Map of residuals in England, Wales, and Scotland in September 2022. White indicates parliamentary constituencies at the expected accident density relative to the power law and population density. Constituencies shaded from orange to red are above expectation, while those shaded from light to dark blue are below expectation. Darker shades represent greater deviations from expectation. Grey areas represent regions with no recorded road accidents for that month.

# 4 Discussions

This study demonstrates consistent rural–urban scaling behaviour in road accident densities across England, Wales, and Scotland over a 14-year period. Rural parliamentary constituencies below the breakpoint generally exhibited sublinear scaling, while urban parliamentary constituencies above it displayed superlinear scaling. These relationships persisted throughout, including during the COVID-19 lockdowns. The pre-exponentials and exponents showed marked seasonality with rural exponents being sub-linear, having summer troughs, and were unchanged over the 14-year period. In comparison, urban regions were consistently super-linear, had winter troughs, and an upward trend. This gradual increase in the urban exponent indicates that the intensity of risk in high density areas has been growing over time, likely driven by congestion, complex traffic flows, and interactions between road users.

To our knowledge, this is the first study to interpret pre-exponential factors in this context. These revealed long-term declines in baseline accident levels, consistent with improvements in road safety, alongside clear seasonal peaks during summer months. Sharp reductions in 2020 and 2021 showed that accidents were sensitive to the national mobility restrictions in place at the time. Fourier-based cross-spectral analysis further demonstrated that pre-exponential factors, rural exponents, and urban exponents are systematically out of phase: rural exponents lagged pre-exponentials by almost five months, while urban exponents were leading by nearly two months. These offsets imply that the drivers of rural and urban accident risks are not only different in scale but also in timing, with rural and urban causes operating on distinct seasonal cycles. Together, scaling exponents and pre-exponential factors provide a fuller understanding of accident scaling: exponents capture how risk responds to changes in population density, while pre-exponential factors offer a descaled, comparable measure of baseline accident density.

Residual analysis highlighted further spatial heterogeneity. Variance was generally stable but spiked during the first lockdown, while skew declined over time, indicating that below expectation accident densities became more common. Modelling residuals with a GLD captured this negative skew and changing variance more effectively than normal assumptions. Mapping residuals revealed systematic negative deviations to the power law in many rural and coastal parliamentary constituencies

These findings underline the need for differentiated policy responses. Analogous to low-emission zones, "low-footprint zones" could be introduced, using the breakpoint as a threshold for stricter regulations or taxation to mitigate disproportionate risks. Such zones would target high-density urban constituencies where scaling exponents indicate accelerating accident risks. Policy measures could include restrictions on the entry of larger or heavier vehicles, dynamic speed regulation, enhanced pedestrian and cyclist protection, or incentives for smaller, lower-risk vehicles. Importantly, the breakpoint provides an empirically grounded criterion for where such interventions would be most effective, ensuring that policy is tailored to environments where risk grows disproportionately with density. Future research should explore how vehicle characteristics (e.g. car size, engine power, weight) and evolving traffic patterns contribute to superlinear scaling behaviour and could inform the design of these low-footprint interventions.

# Data availability statement

Data used in this study are available in the supplementary material.

# Code availability statement

R code for the analysis and the presentation is available in the supplementary material.

# References

1. World Health Organisation. *Global Status Report on Road Safety 2023*. (World Health Organization, 2023).

2. Department for Transport. Reported road casualties in Great Britain, provisional estimates: 2024. https://www.gov.uk/government/statistics/reported-road-casualties-great-britain-provisional-results-2024/reported-road-casualties-in-great-britain-provisional-estimates-2024 (2025).

3. Ritchie, H., Samborska, V. & Roser, M. Urbanization. Our World in Data. https://ourworldindata.org/urbanization (2018).

4. Cabrera-Arnau, C., Prieto Curiel, R. & Bishop, S. R. Uncovering the behaviour of road accidents in urban areas. *R Soc Open Sci* **7**, 191739 (2020).

5. Ghandour, A. J. *et al.* Allometric scaling of road accidents using social media crowd-sourced data. *Physica A: Statistical Mechanics and its Applications* **545**, 123534 (2020).

6. Eksler, V. & Lassarre, S. Evolution of road risk disparities at small-scale level: Example of Belgium. *J Safety Res* **39**, 417–427 (2008).

7. Bettencourt, L. M. A., Lobo, J., Helbing, D., Kühnert, C. & West, G. B. Growth, innovation, scaling, and the pace of life in cities. *Proceedings of the National Academy of Sciences* **104**, 7301–7306 (2007).

8. Bettencourt, L. M. A. The Origins of Scaling in Cities. *Science (1979)* **340**, 1438–1441 (2013).

9. Bettencourt, L. M. A. & West, G. A unified theory of urban living. *Nature* **467**, 913 (2010).

10. Bettencourt, L. M. A., Lobo, J., Strumsky, D. & West, G. B. Urban Scaling and Its Deviations: Revealing the Structure of Wealth, Innovation and Crime across Cities. *PLoS One* **5**, e13541 (2010).

11. Nordbeck, S. Urban Allometric Growth. *Geogr Ann Ser B* **53**, 54–67 (1971).

12. Sutton, J., Shahtahmassebi, G., Ribeiro, H. V. & Hanley, Q. S. Population density and spreading of COVID-19 in England and Wales. *PLoS One* **17**, e0261725 (2022).

13. Sutton, J., Shahtahmassebi, G., Hanley, Q. S. & Ribeiro, H. V. A Heteroscedastic Bayesian Generalized Logistic Regression Model with Application to Scaling Problems. *arXiv preprint arXiv: 2311.17808* https://doi.org/https://arxiv.org/abs/2311.17808 (2023) doi:https://arxiv.org/abs/2311.17808.

14. Hanley, Q. S., Lewis, D. & Ribeiro, H. V. Rural to Urban Population Density Scaling of Crime and Property Transactions in English and Welsh Parliamentary Constituencies. *PLoS One* **11**, e0149546 (2016).

15. Sutton, J., Shahtahmassebi, G., Ribeiro, H. V. & Hanley, Q. S. Rural–urban scaling of age, mortality, crime and property reveals a loss of expected self-similar behaviour. *Sci Rep* **10**, 16863 (2020).

16. Ribeiro, H. V., Hanley, Q. S. & Lewis, D. Unveiling relationships between crime and property in England and Wales via density scale-adjusted metrics and network tools. *PLoS One* **13**, e0192931 (2018).

17. Ribeiro, H. V., Sutton, J. & Hanley, Q. S. Density scaling laws and rural-to-urban transitions. in *Urban Scaling* 98–110 (Routledge, 2024).

18. Sutton, J. *et al.* Comprehensive indicators and fine granularity refine density scaling laws in rural-urban systems. (2025).

19. Finance, O. & Cottineau, C. Are the absent always wrong? Dealing with zero values in urban scaling. *Environ Plan B Urban Anal City Sci* **46**, 1663–1677 (2019).

20. R Core Team. R: A Language and Environment for Statistical Computing. 275–286 Preprint at https://www.r-project.org/ (2013).

21. Chan, C., Chan, G. C. H., Leeper, T. J. & Becker, J. rio: A Swiss-army knife for data file I/O. Preprint at https://rdrr.io/cran/rio/man/rio.html (2021).

22. Dragulescu, A. A. & Arendt, C. xlsx: Read, Write, Format Excel 2007 and Excel 97/2000/XP/2003. Preprint at https://cran.r-project.org/package=xlsx (2018).

23. Pebesma, E. sf: Simple Features for R. *CRAN: Contributed Packages* Preprint at https://doi.org/10.32614/CRAN.package.sf (2016).

24. Warnes, G. R. *et al.* gplots: Various R Programming Tools for Plotting Data. *CRAN: Contributed Packages* Preprint at https://doi.org/10.32614/CRAN.package.gplots (2022).

25. Wickham, H. *et al.* ggplot2: Create Elegant Data Visualisations Using the Grammar of Graphics. *CRAN: Contributed Packages* Preprint at https://doi.org/10.32614/CRAN.package.ggplot2 (2024).

26. Neuwirth, E. RColorBrewer: ColorBrewer Palettes. *CRAN: Contributed Packages* Preprint at https://doi.org/10.32614/CRAN.package.RColorBrewer (2002).

27. Muggeo, V. M. R. segmented: Regression Models with Break-Points / Change-Points Estimation (with Possibly Random Effects). *CRAN: Contributed Packages* Preprint at https://doi.org/10.32614/CRAN.package.segmented (2003).

28. Zeileis, A. & Windberger, T. glogis: Fitting and Testing Generalized Logistic Distributions. *CRAN: Contributed Packages* Preprint at https://doi.org/10.32614/CRAN.package.glogis (2011).